\documentclass[aps,twocolumn,preprintnumbers,amsmath,amssymb]{revtex4}

\usepackage{graphicx}
\usepackage{dcolumn}
\usepackage{bm}
\usepackage{multirow}
\usepackage{color}
\usepackage[colorlinks=true]{hyperref} 
\usepackage{amsmath}
\usepackage{amssymb}
\usepackage{epsfig}
\usepackage[T2A]{fontenc}
\usepackage[cp1251]{inputenc}
\usepackage[russian,english]{babel}
\usepackage{array}

\newcommand{\commentMael}[1]{}

\newcommand{\I}{\left \langle I_z \right \rangle}

\newcommand{\ket}[1]{\left|#1\right\rangle}

\begin{document}

\title{Hyperfine coupling of hole and nuclear spins in symmetric GaAs quantum dots
}

\author{M.~Vidal$^1$}
\author{M.\,V.\,Durnev$^2$}
\author{L.~Bouet$^1$}
\author{T.~Amand$^1$}
\author{M.\,M.\,Glazov$^2$}
\author{E.\,L.\,Ivchenko$^2$}
\author{P.~Zhou$^1$}
\author{G.~Wang$^1$}
\author{T.~Mano$^3$}
\author{T.~Kuroda$^3$}
\author{X.~Marie$^1$}
\author{K.~Sakoda$^3$}
\author{B.~Urbaszek$^1$}

\affiliation{$^1$Universit\'e de Toulouse, INSA-CNRS-UPS, LPCNO, 135 Avenue Rangueil, 31077 Toulouse, France}
\affiliation{$^2$Ioffe Institute, 194021 St.\,Petersburg, Russia}
\affiliation{$^3$National Institute for Material Science, Namiki 1-1, Tsukuba 305-0044, Japan}


\begin{abstract}
In self assembled III-V semiconductor quantum dots, valence holes have longer spin coherence times than the conduction electrons, due to their weaker coupling to nuclear spin bath fluctuations. Prolonging  hole spin stability relies on a better understanding of the hole to nuclear spin hyperfine coupling which we address both in experiment and theory in the symmetric (111) GaAs/AlGaAs droplet dots. In magnetic fields applied along the growth axis,  we create a strong nuclear spin polarization detected through the positively charged trion X$^+$ Zeeman and Overhauser splittings.
The observation of four clearly resolved photoluminescence lines -- a unique property of the (111) nanosystems -- allows us to measure separately the electron and hole contribution to the Overhauser shift. The hyperfine interaction for holes is found to be about five times weaker than that for electrons. Our theory shows that this ratio depends not only on intrinsic material properties but also on the dot shape and carrier confinement through the heavy-hole mixing, an opportunity for engineering the hole-nuclear spin interaction by tuning dot size and shape.
\end{abstract}


\maketitle

\textit{Introduction.---} The spin coherence of carriers in semiconductor quantum dots (QDs) is limited by interactions with the nuclear spin bath fluctuations because the carrier
and nuclear spins are efficiently coupled through the hyperfine interaction \cite{Merkulov:2002a,Urbaszek:2013a,Bechtold:2015a}. In III-V semiconductors the strength of the hyperfine interaction for valence holes is about one order of magnitude weaker than that for the conduction electrons due to the different symmetry of the 
Bloch amplitudes~\cite{grncharova1977, Fischer:2008a}. This results in longer hole spin coherence times and makes the hole spin in QDs an excellent candidate for quantum state manipulations \cite{Eble:2009a,Greilich:2011a,Godden:2012a,DeGreve:2011a,Brunner:2009a,Delteil:2015a}. Further engineering of the hole hyperfine interaction is needed to achieve even longer spin coherence times \cite{Chekhovich:2013a}, both for epitaxially grown and gate defined quantum dots that confine single holes \cite{Tracy:2014a}. However, the hole hyperfine coupling is experimentally difficult to access, the challenge is to separate spectroscopically the weaker hole-nuclear spin interaction from the stronger electron hyperfine interaction \cite{Klochan:2015a}.
Measurements of heavy hole hyperfine constants for holes were performed on QD ensembles \cite{Eble:2009a,Desfonds:2010a,Kurtze:2012a} and for individual QDs in several material systems, namely, InP/GaInP~\cite{Chekhovich:2011a,Chekhovich:2013a} and InGaAs/AlGaAs~QDs~\cite{Fallahi:2010a,Chekhovich:2013a}, as well as in GaAs/AlGaAs interface fluctuation dots \cite{Chekhovich:2013a}. In these experiments, the hole hyperfine coupling constant was measured only in particular QDs with very low point symmetry using ``forbidden'' transitions to detect the Overhauser splittings of dark exciton states. 

\begin{figure*}
\includegraphics[width=0.8\linewidth]{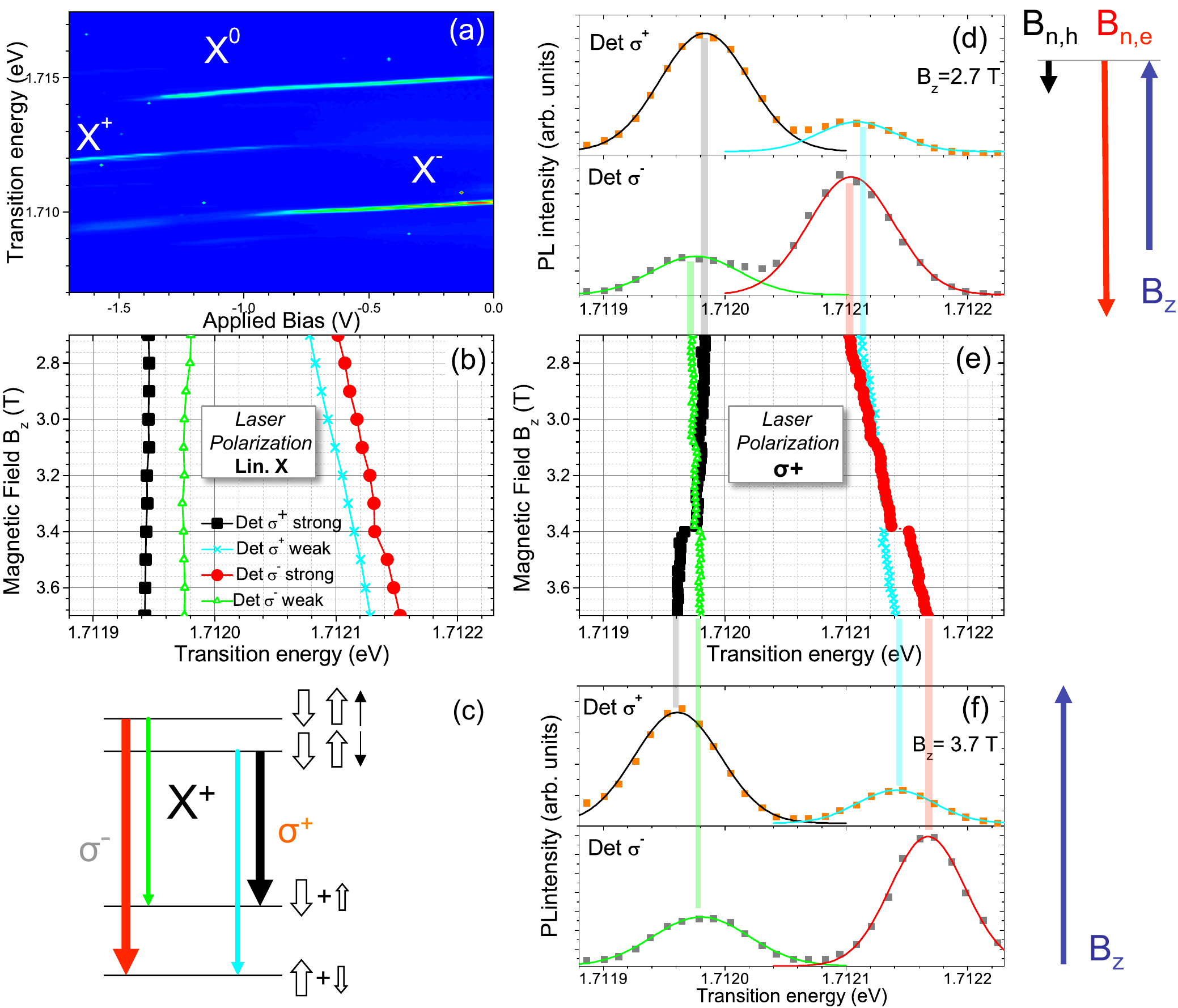}
\caption{\label{fig:fig1} (a) Contour plot of the single dot PL at $T=4$~K as a function of the applied bias voltage using a HeNe laser for above barrier excitation. The neutral exciton X$^0$ and charged trions X$^+$ and X$^-$ are indicated. Blue means less than 100 counts, red $>15000$ counts. (b) PL emission energy of 4 transitions resulting from the X$^+$ recombination in the Faraday geometry $B_z || [111]$ using linearly polarized laser excitation, i.e. avoiding the DNP. (c) Illustration of four optical transitions. Thick and thin long arrows represent strong and weak transitions, respectively; $\sigma^+$ and $\sigma^-$ indicate the circular polarization of emitted photons. The initial state is a three-particle complex X$^+$ with antiparallel hole spins (thick short arrows $\Uparrow, \Downarrow$) and an unpaired electron spin (thin short arrows $\uparrow, \downarrow$); the final states are single heavy-hole admixtures of $\pm 3/2$ spins. (d) PL spectra for the two detection polarizations, $\sigma^+$ and $\sigma^-$, for $B_z$=2.7~T and $B_n\neq$ 0 under the $\sigma^+$ polarized laser excitation. (e) Transition energies as a function of the magnetic field with the circularly polarized laser excitation, i.e., in the presence of the DNP, $B_n\neq 0$, for the four X$^+$ transitions. (f) Same as (d) but for $B_z=3.7$~T and $B_n\simeq 0$. }
\end{figure*}

Here we propose a new approach to measure and eventually engineer the hyperfine interaction for the heavy hole spins in QDs. As a model system we use the strain free GaAs/AlGaAs QDs grown by the droplet epitaxy on (111)A substrates, in contrast to conventional [001]-grown samples used in all the previous studies on the hole hyperfine interaction. As a result of the specific trigonal C$_{3v}$ symmetry of the dots four different exciton states are observed in the magneto-photoluminescence (magneto-PL) as well-separated lines~\cite{Sallen:2011a,Durnev:2013a} of the positively charged trion X$^+$, a complex formed of two holes with antiparallel spins and an unpaired electon, in a magnetic field applied along the growth axis [111], see Fig.~\ref{fig:fig1}. The presence of four lines allows one to separately measure the electron and hole Overhauser shifts for every QD in our system. For the investigated dots we extract a ratio of the heavy hole to electron hyperfine interaction constants $C_{\text{eff}}/A\approx0.2$, which is different in both amplitude and sign compared to the recent studies on [001]-grown GaAs dots~\cite{Chekhovich:2013a}. For the standard (001) QDs the heavy hole hyperfine coupling with a given nucleus is determined by a single constant, mainly determined by the Bloch amplitude~\cite{Fischer:2008a,Testelin:2009a,Chekhovich:2013a}. As we theoretically show, this is not the case for the (111) QDs: Here the heavy hole hyperfine interaction is governed by two constants and the effective Overhauser splitting is strongly affected by the envelope function. Hence, the hyperfine interaction for holes can be engineered by growing dots with different size and shape \cite{Jo:2012a}. An additional advantage of (111) QDs is very strong dynamic nuclear polarization (DNP) \cite{Sallen:2014a} in this strain free system \cite{Stockill:2015a}: In particular, the induced Overhauser splitting of the electron spin sublevels largely exceeds their Zeeman splitting.

\begin{figure*}
\includegraphics[width=0.85\linewidth]{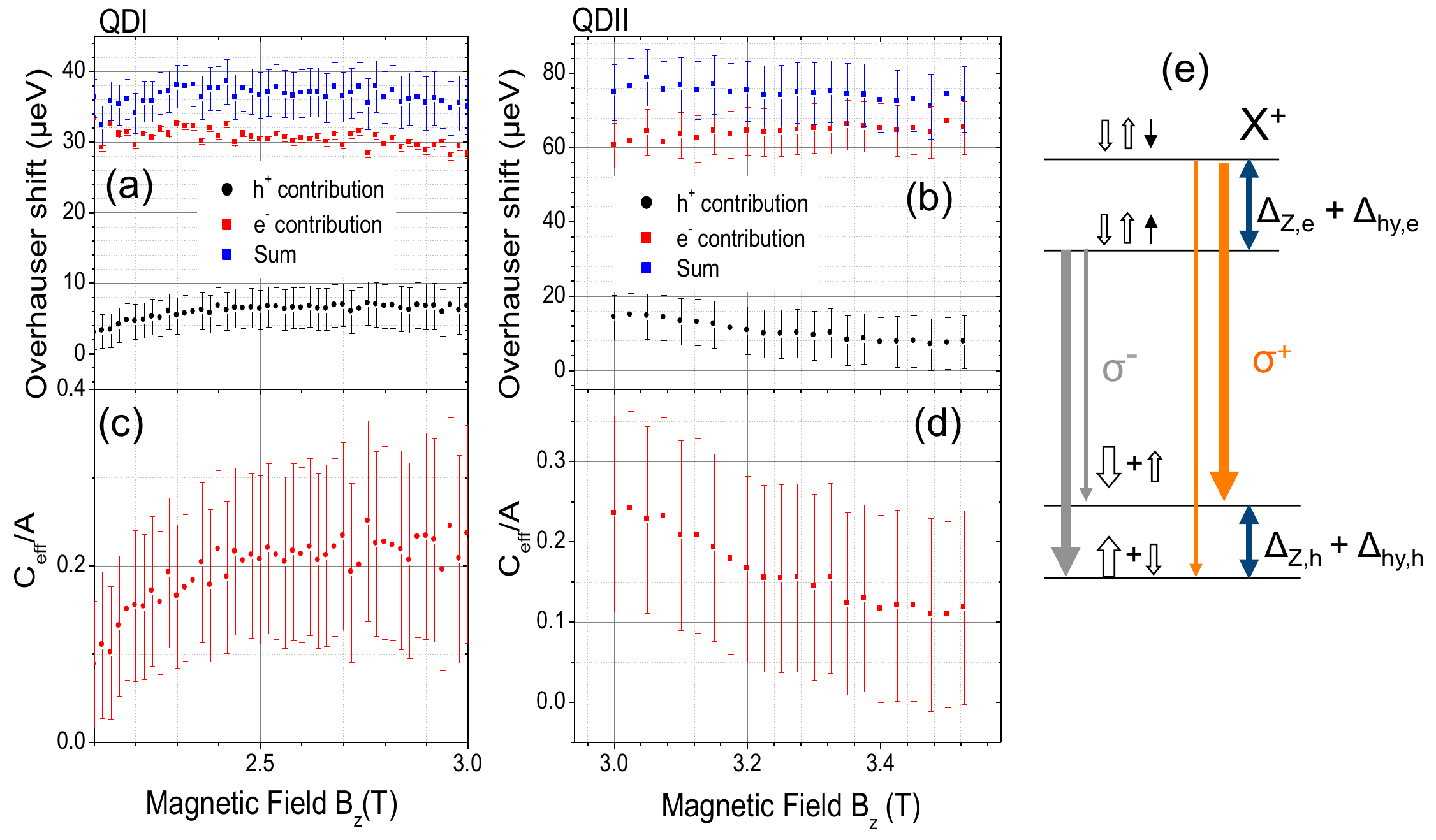}
\caption{\label{fig:fig2} (a) and (b) Overhauser shift for hole (black) and electron (red) and the sum of the two contributions (blue) for two different dots. (c) and (d) the ratio of the effective hole hyperfine constant and electron hyperfine constant. (e) Initial and final states for radiative recombination of X$^+$ in a longitudinal applied magnetic field $B_z$ in the presence of DNP. $\uparrow(\downarrow)$ and $\Uparrow(\Downarrow)$ represent the electron and hole spins, respectively. }
\end{figure*} 

\textit{Samples and Experimental Set-up.---} The sample we study is the same as in Ref.~\onlinecite{Bouet:2014a}. It was grown by droplet epitaxy using a standard molecular beam epitaxy system \cite{Mano:2010a,Sallen:2011a,Belhadj:2008a,Belhadj:2010a}. 
Starting from the $n^+$-GaAs(111)A substrate, the sample consists of 50-nm $n$-GaAs (Si: $10^{18}$~cm$^{-3}$), 100-nm $n$-Al$_{0.3}$Ga$_{0.7}$As (Si: $10^{18}$~cm$^{-3}$), 20-nm Al$_{0.3}$Ga$_{0.7}$As tunnel barrier, GaAs QDs, 120-nm Al$_{0.3}$Ga$_{0.7}$As, 70-nm Al$_{0.5}$Ga$_{0.5}$As, and 10-nm GaAs cap.  In this model system dots are truly isolated as they are not connected by a 2D wetting layer \cite{Mano:2010a,Sallen:2014a}, contrary to Stranski-Krastanov dots and dots formed by quantum well interface fluctuations \cite{Bracker:2005a}. A semitransparent Ti/Au layer with a nominal thickness of 6 nm serves as a Schottky top gate.\\
\indent The single dot PL is recorded at T=4~K with a home build confocal microscope with a detection spot diameter of $\simeq 1$~$\mu$m \cite{Durnev:2013a,Sallen:2014a}. The detected PL signal is dispersed by a double spectrometer and detected by a Si-CCD
camera with the spectral precision of 1~$\mu$eV. Nonresonant optical excitation used for initial dot characterization is achieved by pumping the AlGaAs barrier with a HeNe laser at 1.96~eV. For intra-dot quasi-resonant excitation typically 1~LO-phonon above the transition energy,  a tunable cw Ti-Sa laser (nominal laser FWHM=40~$\mu$eV) is used. The laser polarization control and PL polarization analysis are performed with Glan-Taylor polarizers and liquid crystal waveplates. Magnetic fields up to 9~T can be applied parallel to the growth axis [111] which is also the angular momentum quantization axis and the light propagation axis (Faraday geometry). All results presented here apply to the positively charged trion X$^+$.\\
\indent \textit{Results and Discussion.---} Using a charge tunable device, one can study independently positively charged trions X$^+$, neutral excitons X$^0$ and negatively charged trions X$^-$, see Fig.~\ref{fig:fig1}a and Refs.~\cite{Bouet:2014a,2015arXiv151108196D} for details. The carrier-nuclear spin interaction is probed through the  X$^+$ radiative recombination with the PL circular polarization $P_{c}=({I_{\sigma+}-I_{\sigma-})}/({I_{\sigma+}+I_{\sigma-})}$ directly proportional to the average spin $z$-component of the unpaired electron, $P_c = - 2S_z$. Here $I_{\sigma+}, I_{\sigma-}$ are the integral intensities of the circularly polarized components of the X$^+$ PL emission. Performing LO-phonon assisted  excitation of the trion ground state with a $\sigma^+$ polarized laser, we observe a highly $\sigma^+$ polarized emission of the X$^+$ trion with  $P_c>80\%$~\cite{supinfo}.  Application of a longitudinal magnetic field $B_z$ $||$ [111] results in four optically allowed emission lines, two lines per each circular polarization \cite{Sallen:2011a,Durnev:2013a}, see scheme in Fig.~\ref{fig:fig1}c. A typical evolution of the four optical transitions with the field $B_z$ is presented for linearly polarized laser excitation in Fig.~\ref{fig:fig1}b.
Under the linearly polarized excitation, no  DNP is achieved and the lines evolve continuously with increasing the applied magnetic field, Fig.~\ref{fig:fig1}b. \\
\indent A completely different behaviour of the PL lines is observed for the circularly polarized excitation which results in optical pumping of carrier and nuclear spins, see Fig.~\ref{fig:fig1}e. First of all, we observe an additional Overhauser shift (OHS) of the PL lines proportional to the effective magnetic field $\bm B_{n} \parallel z$ experienced by the carriers due to the nuclear spin polarization \cite{Gammon:1997a}. Secondly, at $B_z =B_{\rm cr}\approx 3.4$~T the order of the four transitions changes abruptly: at $B_z < B_{\rm cr}$ the outer spectral lines are substantially weaker than the inner lines (Fig.~\ref{fig:fig1}d), whereas for $B_z > B_{\rm cr}$ the strong intensity lines become the outer ones (Fig.~\ref{fig:fig1}f), as in the case of linearly polarized laser excitation. The abrupt jumps of the PL lines at $B_z=B_{\rm cr}$ indicate the nuclear spin polarization collapse and inefficiency of DNP. This is a direct signature of the non-linear carrier-nuclear spin interaction observed already in other QD systems \cite{Eble:2006a,Maletinsky:2007b,Tartakovskii:2007a,Krebs:2010a,Urbaszek:2013a} and illustrated in more detail in the supplement~\cite{supinfo}.
Moreover, according to Fig.~\ref{fig:fig1}d, at $B_z<B_{\rm cr}$ the effective nuclear field $B_{n,e}$ acting on the electron spin is larger in magnitude and opposite in sign to $B_z$. As a result, the order of electron spin-up and spin-down levels is \textit{reversed} compared to the case where the DNP is absent, Figs.~\ref{fig:fig1}b,f. At $B_z=B_{\rm cr}$ the DNP collapses and the normal order of the four optically allowed transitions is restored. The critical field $B_{\rm cr}$ for the nuclear polarization collapse varies from dot to dot and, in this nanosystem, takes values between 2 T and 4~T.\\
\indent Next we show that the hole hyperfine interaction constant can be measured by analysing the four well separated lines due to the X$^+$ optical recombination. We recall that the four emission lines rather than two appear in magneto-PL from magneto-induced heavy-hole mixing, the effect related with the trigonal, $C_{3v}$, point symmetry of (111) QDs~\cite{Sallen:2011a}. The Zeeman effect for heavy holes is thus governed by two $g$-factors, $g_{h1}$ and $g_{h2}$, describing the diagonal and off-diagonal contributions to the effective Hamiltonian in the basis of $|\pm 3/2\rangle$ states,
\begin{equation}
\label{HB}
\mathcal H_B = \frac{1}{2} \mu_B B_z {\left( g_{h1} \sigma_{z}^{(h)} + g_{h2} \sigma_{x}^{(h)} \right)}\:,
\end{equation} 
where $\sigma_x^{(h)}$ and $\sigma_z^{(h)}$ are the effective heavy-hole spin matrices~\cite{Sallen:2011a,Durnev:2013a}. Due to nonzero value of $g_{h2}$, the hole states are mixtures of the $|\pm 3/2\rangle$ basic states enabling four radiative recombination channels for the X$^+$ trion, see Fig.~\ref{fig:fig1}c. 
As a result, the QD symmetry provides a natural optical access to all four exciton states and allows us to measure precisely the separate heavy-hole and electron contributions to the Overhauser shifts. In contrast, the [001]-grown dots with the $C_{2v}$ symmetry show only two emission lines for excitons in a magnetic field applied in the Faraday geometry \cite{Bayer:2000a,Bayer:2002a}. If the symmetry of [001]-grown dots is reduced to C$_s$ or
C$_2$, e.g., due to a QD distortion or an in-plane strain, two extra lines
appear allowing one to separate the hole and electron Overhauser shifts in these very particular dots, as in Refs.~\cite{Chekhovich:2011a,Chekhovich:2013a}. On the other hand, a deviation from the C$_{3v}$ symmetry of many [111]-grown dots in our samples is negligible and they are excellent sources of entangled photon pairs (biexciton-exciton cascade) \cite{Kuroda:2013a}. The four transitions that we use here are all optically allowed for perfectly symmetric $C_{3v}$ dots and therefore the hole and electron Overhauser shift determination does not rely on finding an exotic dot with reduced symmetry. 
Further  details of the electron and hole OHS extraction are presented in the Supplementary information \cite{supinfo}. Particularly, the hole contribution to the OHS is clearly seen for the two typical dots from different samples, see Fig.~\ref{fig:fig2}. In the presence of DNP the electrons and holes experience Overhauser shifts quantified as additional contributions  $\text{$\Delta$}_{\text{hy},e} =A \left\langle I_z\right\rangle$ and $\text{$\Delta $}_{\text{hy},h}=C_{\text{eff}} \left\langle I_z\right\rangle$ to the electron and hole Zeeman splittings, respectively, Fig.~\ref{fig:fig2}(e). Here $\left\langle I_z\right\rangle$ is the average nuclear spin projection on the $z$-axis~\cite{Urbaszek:2013a}, $A$ and $C_{\rm eff}$ are the effective electron and hole hyperfine coupling constants. The experiment allows us to extract the ratio $C_{\text{eff}}/A \approx 20$\% for typical dots, see Fig.~\ref{fig:fig2}, which is the average value taken from measurements in the magnetic field range where the Overhauser field is fairly constant. An eventual weak dependence of the coefficient $C_{\text{eff}}$ on $B_z$ is discussed in the supplement~\cite{supinfo}. Surprisingly, our measured ratio $C_{\text{eff}}/A \approx 20$\% is about four times larger than the value reported for GaAs/AlGaAs interface fluctuation dots \cite{Chekhovich:2013a} and, moreover, in InGaAs dots the values of about one tenth with the opposite sign were reported~\cite{Chekhovich:2013a,Fallahi:2010a}.\\
\indent This unexpected result is due to the C$_{3v}$ point symmetry of the (111) dots which dictates the following form of the hole-nuclei interaction Hamiltonian~\cite{supinfo}
\begin{multline}
\label{eq:hh_hyper}
\mathcal H_{\mathrm{hy}, h} =  {\left( C_1 \sigma_z^{(h)} +  C_2 \sigma_x^{(h)} \right)} \frac{\langle I_z \rangle}{2}
 = \begin{pmatrix}
 C_{1} &  C_{2}\\
 C_{2} & - C_{1}
\end{pmatrix}  \frac{\langle I_z \rangle}{2}\:.
\end{multline}
Like the Zeeman Hamiltonian \eqref{HB}, the hyperfine interaction for holes is described by two constants, $C_1$ and $C_2$, the latter being responsible for the heavy-hole mixing induced by the Overhauser field.

For sufficiently strong external magnetic fields such as $\sqrt{C_1^2 + C_2^2} \langle I_z \rangle \ll g_h \mu_B B_z$, where $g_h=\sqrt{g_{h1}^2+g_{h2}^2}$, the heavy-hole spin states are defined by the Zeeman effect, and the hyperfine Hamiltonian~\eqref{eq:hh_hyper} can be treated as a small perturbation with respect to Eq.~\eqref{HB}. In this case one has
\begin{equation}
\label{eq:Cinf}
C_{\rm eff} = \frac{g_{h1}}{g_h} C_1 + \frac{g_{h2}}{g_h} C_2\:.
\end{equation} 
It follows then that the heavy-hole hyperfine interaction in the dots under study is defined not only by the material constants $C_1$ and $C_2$ but also, through the effective $g$-factors $g_{h1}$ and $g_{h2}$, by the dot size and shape. A possible origin of $C_2$ can be the cubic anisotropy contributions to the hyperfine coupling~\cite{Chekhovich:2013a} as well as orbital contributions caused by an admixture of light-hole to  heavy-hole states (further analysis is beyond the scope of this paper)~\cite{supinfo}. We stress that this is a major difference compared to the previous studies on hole-nuclear spin interaction which used the [001]-grown QDs, where the size and shape of the QD only weakly affects the $C_{\rm eff}/A$ ratio.\\
\indent \textit{Conclusion.---} In summary, in the symmetric [111]-grown GaAs droplet dots we measure, without the need in symmetry breaking, the hole hyperfine constant  to be $\sim$5 times smaller than the electron one. Using symmetry considerations we have shown that the sign and amplitude of the hole hyperfine interaction constant in the dots of the $C_{3v}$ symmetry reveal an important advantage as compared to the standard (001) QD systems investigated so far. 
The hole hyperfine constant in the (111) quantum dots is sensitive to the dot geometry through the light-heavy-hole mixing opening the way for engineering the hole-nuclear spin interaction~\cite{Jo:2012a,Bree:2016a}. 

We acknowledge funding from ERC Grant No. 306719 and LIA CNRS---Ioffe RAS ILNACS. M.V.D., M.M.G, and E.L.I. were supported in part by the Russian Science Foundation, project \#14-12-01067, RF president grants MD-5726.2015.2, MK-7389.2016.2, and the  ``Dynasty'' foundation . X.M. acknowledges funding from the Institut Universitaire de France.

\newpage

\section*{SUPPLEMENTARY INFORMATION}

\section{Efficient optical pumping of nuclear spins in (111) $\text{GaAs}$ quantum dots}\label{sec:moreexp}

Experimentally, for optical orientation of charge-carrier and nuclear spins we use quasi-resonant, circularly polarized excitation. In our charge tunable device, we select a bias region where the positively charged trion X$^+$ emission dominates. As the two holes form a spin singlet, the average polarization $P_c$ of the X$^+$ PL emission is determined by the electron spin projection $S_z$ as $P_c=-2S_z$. This allows us to measure the electron spin polarization via the circular polarization degree of the PL and the nuclear spin polarization via the Overhauser shift~\cite{Eble:2006a}. If we plot both quantities as a function of the applied magnetic field $B_z$ we see the bistable behaviour typical of the charge carrier spin system coupled to the nuclear spins via the hyperfine interaction \cite{Eble:2006a,Maletinsky:2007b,Tartakovskii:2007a,Krebs:2010a,Urbaszek:2013a}. In Fig.~1 of the main paper we only plot the sweep with increasing field direction, which shows the dynamic nuclear polarization (DNP) collapse at $B_z=3.4~T$. The measurements of the hole and electron contribution to the hyperfine interaction are carried out well before the collapse, at magnetic fields $B_z$ around 2~T in Fig.~2 of the main text.

\begin{figure}[b]
\includegraphics[width=0.95\linewidth]{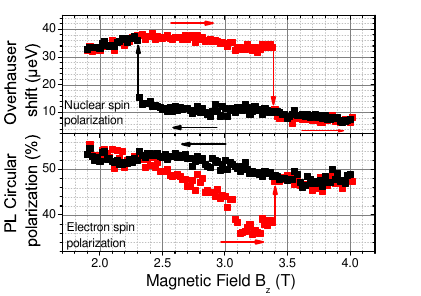}
\caption{\label{fig:figS1} Upper panel: The total Overhauser shift (electron + hole contribution) is plotted as a function of magnetic field $B_z$ for different sweep directions. Lower Panel: The PL polarization of the X$^+$ emission is plotted, showing a bistability of the electron spin polarization as $P_c=-2S_z$.}
\end{figure}

\section{Heavy-hole and electron hyperfine Hamiltonians in (111) quantum dots}\label{sec:hyperfine}

\subsection{General symmetry analysis}

We recall that in the (111) QDs with the $C_{3v}$ point symmetry the electron Kramers doublet transforms according to the two-dimensional irreducible representation $\Gamma_4$, while the heavy-hole doublet transforms according to the reducible representation $\Gamma_5\oplus \Gamma_6$~\cite{koster63}. As a result, the effective Hamiltonians of hyperfine interaction can be conveniently expressed via the corresponding electron spin-1/2 and hole-pseudospin-1/2 operators represented by the sets of the Pauli matrices $\bm \sigma^{(e)} = (\sigma^{(e)}_x, \sigma^{(e)}_y, \sigma^{(e)}_z)$, $\bm \sigma^{(h)} = (\sigma^{(h)}_x, \sigma^{(h)}_y, \sigma^{(h)}_z)$. Hereafter we use the frame $x\parallel [11\bar 2]$, $y\parallel [\bar 110]$, $z\parallel [111]$ related to the growth axis of the dots.

The Ga and As isotopes have nuclear spins $I=3/2$. Therefore, nuclear states transform according to the reducible representation $\Gamma_4\oplus\Gamma_5 \oplus \Gamma_6$. The splittings of nuclear states with $|I_z|=1/2$ and $|I_z|=3/2$ are controlled by quadrupolar interactions, which are expected to be minor in our strain-free QDs. 
Hence, it is convenient to describe the nuclear spin as a pseudovector $\bm I = (I_x, I_y, I_z)$ composed of the angular momentum 3/2 matrices and treat quadrupolar effects for nuclei separately. By contrast, in our QDs the splitting between the heavy- and light-hole states is substantial, $\gtrsim 10$~meV~\cite{Durnev:2013a,Bouet:2014a}, therefore it suffices to focus on the heavy-hole doublet only. Table~\ref{tab:table1} presents the combinations of basic matrices of angular momentum $3/2$ transforming according to the irreducible representations of the $C_{3v}$ point group. Additionally, we marked the combinations in accordance to their parity under the time-reversal.

\begin{table}[h]
\caption{\label{tab:table1}  Combinations of basic matrices $I=3/2$ transforming according to the irreducible representations of $C_{3v}$ point symmetry group, $z\parallel C_3$, $\sigma_v$ is the $(xz)$-plane.}
\begin{ruledtabular}
\begin{tabular}{c|c|c}
Representation & Time reversal & Basic matrices\\
\hline
\multirow{ 2}{*}{$\Gamma_1$~(identity)} & even & $I$ (unit $4\times 4$); $I_z^2$;\\
& odd & $I_y^3-3\{I_y,I_x^2\}_s$;\\
\hline
$\Gamma_2$ & odd & $I_z$; $I_x^3-3\{I_x,I_y^2\}_s$; $I_z^3$;\\
\hline
\multirow{ 3}{*}{$\Gamma_3$~(E; pseudovector)} & \multirow{2}{*}{odd} & $(I_x,I_y)$; $(\{I_x,I_z^2\}_s,\{I_y,I_z^2\}_s)$; \\
& & $(\{I_x^2-I_y^2,I_z\}_s,-2\{I_xI_yI_z\}_s)$; \\
\cline{2-3}
& \multirow{2}{*}{even} & {$(2\{I_y,I_x\}_s,I_x^2-I_y^2,)$};\\
& &  $(\{I_y,I_z\}_s,-\{I_x,I_z\}_s)$.
\end{tabular}
\end{ruledtabular}
\end{table}

It follows then that, in the C$_{3v}$ point group, the matrices $\sigma^{(e)}_z$, $I_z$, $I_z^3$,  and $I_x^3 - 3 \{ I_x, I_y^2 \}_s$, where $\{ \ldots \}_s$ denote symmetrized product, transform according to the $\Gamma_2$ irreducible representation {and are odd under the time reversal}. The {matrices $(\sigma_x^{(e)}, \sigma_y^{(e)})$ form the basis of two-dimensional representation $\Gamma_3$ and are also odd with respect to the time inversion. {Making use of Tab.~\ref{tab:table1} we derive} the hyperfine Hamiltonian for an electron ($\mathcal H_{{\rm hy},e}$) {with a single nucleus parametrized by the six coefficients $A_\parallel$, $A_\parallel'$, $A_\parallel''$, $A_\perp$, $A_\perp'$, $A_\perp''$} and has the following form
\begin{multline}
\label{eq:H_hy_e}
\mathcal H_{{\rm hy},e} = \frac{A_{{\parallel}}}{2} \sigma_z^{(e)} I_z + \frac{A_{{\perp}}}{2} \left( \sigma_x^{(e)} I_x + \sigma_y^{(e)} I_y \right) \\
+  {\frac{A_{\parallel}'}{2} \sigma_z^{(e)} I_z^3}+ \frac{{A_{\parallel}''}}{2} \sigma_z^{(e)} \left( I_x^3 - 3\{ I_x, I_y^2 \}_s \right)\\
 + {\frac{A_{\perp}'}{2} \left\{ \sigma_x^{(e)} I_x + \sigma_y^{(e)} I_y ,I_z^2\right\}_s} \\
 + {\frac{A_{\perp}''}{2} \left\{ \sigma_x^{(e)}(I_x^2- I_y^2) - 2\sigma_y^{(e)} \{I_xI_y\}_s ,{I_z}\right\}_s}\:.
\end{multline}
In the isotropic approximation $A_\parallel = A_\perp$ and other parameters vanish. Note that, in general, the bulk hyperfine  Hamiltonian for the electron has the form $\bm \sigma^{(e)}\cdot(A \bm I  + A'\bm I^3 )/2$ described by two constants, with the $\bm I^3=(I_{x'}^3,I_{y'}^3,I_{z'}^3)$ in the cubic axes frame $x'$, $y'$, $z'$. The second contribution nonlinear in $\bm I$ is small and usually disregarded.
 The terms with $A_\parallel$, $A_\perp$, $A_\parallel'$ and $A_\perp'$ are allowed in the uniaxial approximation. The remaining terms (with $A_\parallel''$ and $A_\perp''$) result from the trigonal symmetry of the quantum dot. These terms describe noncollinear coupling of the electron and nuclear spins, which may result, e.g., in the nuclear spin relaxation and ``dragging'' of QD transition with the laser at a resonant excitation~\cite{Urbaszek:2013a}. 
Note, that the light-hole hyperfine interaction is described by the Hamiltonian of the same form as Eq.~\eqref{eq:H_hy_e} because the light hole states transform according to the two-dimensional irreducible representation~$\Gamma_4$.

{For heavy holes, the basic matrices $\sigma^{(h)}_z$ and $\sigma^{(h)}_x$ transform according to the $\Gamma_2$ irreducible representation of the $C_{3v}$ point group, while the matrix $\sigma^{(h)}_y$ is invariant under all point-group transformations and is odd at time reversal. As a result,} the effective heavy-hole {hyperfine coupling} Hamiltonian{,} $\mathcal H_{{\rm hy}, h}${,} assumes the form
\begin{multline}
\label{eq:H_hy_h}
\mathcal H_{{\rm hy}, h} = \frac12 \left( C_1 \sigma_z^{(h)} + C_2 \sigma_x^{(h)} \right) I_z + \frac12 \left( C_3 \sigma_z^{(h)} + C_4 \sigma_x^{(h)} \right) I_z^3\\+ \frac12 \left( C_5 \sigma_z^{(h)} + C_6 \sigma_x^{(h)} \right)\left( I_x^3 - 3\{ I_x, I_y^2 \}_s  \right) \\
+\frac12 C_7 \sigma_y^{(h)} (I_y^3-3\{I_y,I_x^2\}_s) \:.
\end{multline}
It is noteworthy that all coefficients $C_l$ ($l=1..7$) except for $C_1$ and $C_3$ are related with the trigonal symmetry of the dot; particularly the terms proportional to $C_2$ and $C_4$ may result in the hyperfine-coupling induced heavy-hole spin relaxation. Noncollinear terms, as corresponding terms in the electron Hamiltonian~\eqref{eq:H_hy_e}, may be derived in the high-order perturbation theory. In the following we focus on the coefficients $C_1$ and $C_2$ involved in the description of the optical orientation of spins along the $z$ axis in the longitudinal magnetic field ${\bm B}$.

{\subsection{Relation with the bulk hyperfine coupling}}

{The hyperfine interaction of the electron (hole) and nuclear spins is dominated by the Bloch functions of the charge carriers in the vicinity of the nucleus~\cite{ll3_eng,Abragam:1961a,perel76eng}. Therefore, it could be expected that the main contributions to the hyperfine interaction Hamiltonians, Eqs.~\eqref{eq:H_hy_e}, \eqref{eq:H_hy_h}, result from the averaging of the bulk Hamiltonian over the charge carriers wavefunctions. In particular, the heavy-hole hyperfine coupling is described in the T$_d$ point symmetry relevant to the bulk GaAs by the general Hamiltonian where we retain only linear in nuclear spin $\bm I$ terms:}
\begin{equation}
\label{eq:H_hy_Td}
\mathcal H_{{\rm hy}, h}^{[001]} = \frac{A}{2} \left(M_1 \bm J\cdot \bm I + M_2 \bm J^3 \cdot \bm I\right)\:, 
\end{equation}
where  we use the  coordinate frame $x'\parallel [001]$, $y'\parallel [010]$, $z'\parallel [001]$ related to the cubic crystallographic axes, and, to shorten the notations, $\bm J^3 = (J_{x'}^3, J_{y'}^3, J_{z'}^3)$,
$A$ is the isotropic hyperfine coupling constant for the electron, $M_1$ and $M_2$ are dimensionless parameters. To derive this Hamiltonian we use the fact that the pseudovectors $\bm J$, $\bm I$, $\bm J^3$  transform according to the same $F_1$ irreducible representation of the T$_d$ point group. 
In the Hamiltonian~\eqref{eq:H_hy_Td} we leave only contributions, which are odd in components of $\bm J$ and $\bm I$ and lead to the splitting of the hole spin sublevels.
In bulk semiconductors, the isotropic part of the hole hyperfine interaction, proportional to the $ M_1$,  has been addressed in a number of works, see, e.g., Refs.~\cite{perel76eng,Fischer:2008a,Testelin:2009a}, microscopically it can be derived taking into account the dipole-dipole interaction of nuclear spins with the hole states assuming atomic-like $p$-shell orbital Bloch functions $\mathcal X(\bm r)$, $\mathcal Y(\bm r)$ and $\mathcal Z(\bm r)$ of the valence band. In Ref.~\cite{Chekhovich:2013a} it has been pointed out that the $T_d$ symmetry allowed admixture of $d$-shell atomic orbitals to $p$-shell in orbital functions yields $M_2\ne 0$.

{Transformation of Eq.~\eqref{eq:H_hy_Td} to the reference frame $x$, $y$, $z$, relevant to the (111) QDs results in the effective Hamiltonian}
\begin{widetext}
\begin{equation}
\label{eq:H111}
\mathcal H_{{\rm hy},h}^{[111]} = \frac{A}{16} \left(
\begin{array}{cccc}
(12M_1 + 23M_2)I_z & \sqrt{3}(4M_1+9M_2)I_- & -2\sqrt{6}M_2 I_+ & -4\sqrt{2} M_2 I_z \\
\sqrt{3}(4M_1+9M_2)I_+ & (4M_1 + 13M_2)I_z & 2(4M_1+7M_2)I_- & 2\sqrt{6} M_2 I_+ \\
-2\sqrt{6}M_2 I_- & 2(4M_1+7M_2)I_+ & -(4M_1 + 13M_2)I_z & \sqrt{3}(4M_1+9M_2) I_- \\
-4\sqrt{2} M_2 I_z & 2\sqrt{6} M_2 I_- & \sqrt{3}(4M_1+9M_2) I_+ & -(12M_1 + 23M_2)I_z
\end{array}
\right)\:{,}
\end{equation}
\end{widetext}
{and} yields the following expressions for the hyperfine constants of the localized hole:
\begin{subequations}
\label{eq:C1C2}
\begin{align}
\frac{ C_1}{  A} &= \frac{1}{2}\left(3M_1 + \frac{23}{4} M_2\right) = \frac{12}{5} \tilde M_p - \frac{24}{7} \tilde M_d, \\
 \frac{ C_2}{  A} &= -\frac{\sqrt{2}}{2} M_2 = - \frac{6\sqrt{2}}{7} \tilde M_d\:,
\end{align}
\end{subequations}
where $\tilde M_p$ and $\tilde M_d$ are the parameters of the microscopic theory of Ref.~\cite{Chekhovich:2013a}, which describe the contributions of $p$- and $d$-symmetry components in the atomic part of the $\Gamma_8$-hole Bloch function.

\subsection{Account for the heavy-light-hole mixing}
The values of $C_1$ and $C_2$ are modified if one takes into account the heavy-light-hole mixing in the studied dots. The two ground spin states of the heavy-hole in the C$_{3v}$ dots can be presented as linear combinations of the $\ket{\pm3/2}$ and $\ket{\pm1/2}$ Bloch functions~\cite{Durnev:2013a}
\begin{equation}
\label{psi:trig}
\Psi_{\pm}^{(hh)} (\bm r) = S(\bm r) \ket{\pm3/2} \pm \alpha P_\pm (\bm r) \ket{\pm 1/2} + \beta {P_\mp'}(\bm r) \ket{\mp 1/2}\:.
\end{equation}
Here $S(\bm r)$ is a function which is invariant under C$_{3v}$ point-group operations (the scalar representation $\Gamma_1$), and {two pairs of functions $[P_+(\bm r), P_-(\bm r)]$, $[P_+'(\bm r), P_-'(\bm r)]$, $P_+=P_-^*$, $P_+'={P_-'}^*$,  form two bases of irreducible representation} $\Gamma_3${. The functions $P_\pm(\bm r)$, $P'_\pm(\bm r)$ are assumed to be normalized to unity,} $\alpha$ and $\beta$ {Eq.~\eqref{psi:trig} are real} dimensionless constants {describing the admixture of the light-hole, $|\Gamma_8,\pm 1/2\rangle$, states to the heavy-hole, $|\Gamma_8,\pm 3/2\rangle$, ones}.

Noteworthy, in the uniaxial approximation $P_\pm\propto \exp{(\pm \mathrm i \varphi)}$ and $P_{\pm}' \propto \exp{(\mp 2\mathrm i \varphi)}$, where $\varphi$ is the azimuthal angle of the position vector $\bm r$. In trigonal QDs, however, the second harmonics transform in the same way as the first ones, and the function pairs $P_\pm$ and $P_\pm'$ have same angular dependence $\propto \exp{(\pm \mathrm i \varphi)},$ $\exp{(\mp2 \mathrm i \varphi)}$.

By projecting Hamiltonian~\eqref{eq:H111} at $I_x = I_y = 0$ onto $\Psi_\pm$ states we obtain the corrected values of the hyperfine constants:
\begin{subequations}
\label{eq:C1C2_corr}
\begin{align}
\frac{\tilde{C}_1}{ A} &= \frac{12}{5} \tilde M_p - \frac{24}{7} \tilde M_d + \frac45 \left( \alpha^2 - \beta^2 \right) \tilde{M}_p, \\
 \frac{\tilde{C}_2}{  A} &=  - \frac{6\sqrt{2}}{7} \tilde M_d + \frac{8}{5} \alpha \beta {Q}\tilde{M}_p \:{,} \label{C2lh}
\end{align}
\end{subequations}
{where the $Q=\int d \bm r P_+(\bm r)P_-'(\bm r) $. For an axially symmetric system, the product $P_+(\bm r) P_-'(\bm r)$ is proportional to $\exp{(3 \mathrm i \varphi)}$ and the integral $Q$ vanishes. Therefore this integral can serve a measure of the QD triangularity. Equations~\eqref{eq:C1C2_corr} demonstrate that in trigonal QDs the heavy-light hole mixing results not only in a renormalization of the ``diagonal'' hyperfine coupling constant $C_1$, the effect known for standard (001) QDs~\cite{Testelin:2009a,Chekhovich:2013a}, but also in an appearance of nonzero $C_2$, even neglecting cubic anisotropy terms in the bulk Hamiltonian~\eqref{eq:H_hy_Td}. We stress that this effect is absent in standard (001) QDs where $Q$ vanishes from the symmetry arguments.}

\section{Heavy-hole Overhauser shift in external magnetic field}

{\subsection{Interplay of Overhauser and Zeeman effects}}\label{sec:interplay}

We will now analyze the heavy-hole Overhauser shift in the presence of external longitudinal magnetic field $\bm B = (0,0,B_z)$. A heavy-hole localized in a C$_{3v}$ dot experiences a joint action of the hyperfine field described by $\mathcal H_{{\rm hy}, h}$ [see Eq.~\eqref{eq:H_hy_h}], and the external magnetic field described by the Zeeman Hamiltonian~\cite{Sallen:2011a}
\begin{equation}
\label{Zeeman}
\mathcal H_{\rm Z} = \frac{1}{2}  \mu_B B_z 
\begin{pmatrix}
g_{h1} & g_{h2}\\
g_{h2} & -g_{h1}
\end{pmatrix}\:.
\end{equation}
Here $g_{h1}$ and $g_{h2}$ are the two effective heavy-hole $g$-factors; the microscopic theory of these parameters is developed in Ref.~\cite{Durnev:2013a}. We note that at at the nonzero magnetic field $B_z$ and average value $\langle I_z \rangle\ne 0$, both the hyperfine ($\mathcal H_{{\rm hy}, h}$) and the Zeeman ($\mathcal H_{\rm Z}$) Hamiltonians couple the heavy holes with the $\pm 3/2$ angular momentum projections onto the $z$-axis. 

The eigenenergies of the total Hamiltonian $\mathcal H_{{\rm hy}, h} + \mathcal H_{\rm Z}$ are given by
\begin{equation}
\label{eq:hole_levels}
\varepsilon_\pm = \pm \frac12 \sqrt{\left( g_{h1} \mu_B B_z + C_1 \I  \right)^2 + \left( g_{h2} \mu_B B_z + C_2 \I \right)^2}\:.
\end{equation}
It is worth to mention that, for the optical transitions from the initial X$^+$ state with the electron spin $s = \pm 1/2$ to the final single-hole state $\pm$, the Zeeman and hyperfine contributions to the photon energy is given by $s \mu_B B_z + \varepsilon_{\mp}$.

We define the Overhauser shift as  the difference:
\begin{equation}
\label{OSS:gen}
\Delta_{\rm OS} = \left(\varepsilon_+ - \varepsilon_- \right) - g_h \mu_B |B_z|
\end{equation}
\begin{widetext}
\[
=\sqrt{\left( g_{h1} \mu_B B_z + C_1 \I  \right)^2 + \left( g_{h2} \mu_B B_z + C_2 \I \right)^2} - g_h\mu_B |B_z|.
\]
\end{widetext}
Here $g_h \mu_B B_z$ is the Zeeman splitting of the hole spin levels, i.e., the splitting at $\I = 0$, $g_h = \sqrt{g_{h1}^2 + g_{h2}^2}$.  Under experimental conditions the Zeeman splitting caused by the external field exceeds by far the Overhauser shift for holes making it possible to reduce Eq.~\eqref{OSS:gen} to
{\begin{equation}
\label{OSS}
\Delta_{\rm OS} = C_{\rm eff} \I, 
\end{equation}}
with
\begin{equation}
\label{Ceff}
C_{\rm eff} = \left(\frac{g_{h1}}{g_{h}}C_1 + \frac{g_{h2}}{g_h}C_2\right)\frac{B_z}{|B_z|}\:{,}
\end{equation}
in agreement with Eq.~(3) of the main text.
In this limit the hole spin states are fixed by the magnetic field and the nuclear effect can be evaluated in the first-order perturbation theory. Interestingly, the effective hyperfine coupling constant depends both on the magnitude and sign of $g_{h2}$.
Note that, contrary to conventional [001]-grown quantum dots, the effective hyperfine constant $C_{\rm eff}$ in the (111) dots depends not only on the material characteristics $C_1$ and $C_2$, but also, through $g_{h1}$ and $g_{h2}$, on the particular dot size and shape.

It is worth to mention that, as it follows from Eq.~\eqref{OSS:gen}, the Overhauser shift is a nonlinear function of the nuclear spin polarization $\langle I_z \rangle$ and depends on the external field $B_z$ for fixed $\langle I_z \rangle$. This results in a weak dependence of $C_{\rm eff}$ defined by Eq.~\eqref{OSS} on the magnetic field as demonstrated in Fig.~2 of the main text.

{\subsection{Numeric estimations of hyperfine constants and comparison with experiments}} \label{sec:numeric}

Let us introduce the effective nuclear fields $B_N^{(e)} = A \langle I_z\rangle/ (g_e \mu_B)$ and
$B_N^{(h)} = {|C_{\rm eff} \langle I_z\rangle|}/ (g_h \mu_B)$ acting on an electron and a hole. Their ratio is given by $B_N^{(h)}/B_N^{(e)} = ({|C_{\rm eff}|}/A) (g_e/g_h)$. The analysis of intensities of lines gives $g_e>0$ and $g_{h1}<0$ in our dots~\cite{durnev2015}, moreover, as follows from Fig.~1(b) of the main text, $g_e/g_h\sim 0.1$. Assuming ${|C_{\rm eff}|}/A = 0.2$ we obtain $B_N^{(h)}/B_N^{(e)} \sim 0.02$. In the studied range of external magnetic fields we have $|B_N^{(e)}| \sim (1\div1.5) B_z$, in agreement with Fig.~1(e) and Fig.~2 in the main text. Therefore,  $|B_N^{(h)}|/B_z \sim 0.1\div0.15$ and $B_z \gg |B_N^{(h)}|$ which is confirmed by comparison of the Zeeman splitting and the Overhauser shift presented in Fig.~\ref{fig:figS2}. This justifies the validity of Eqs. \eqref{OSS} and \eqref{Ceff}.

\begin{figure}
\includegraphics[width=0.95\linewidth]{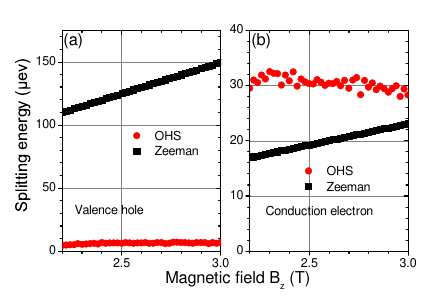}
\caption{\label{fig:figS2} The measured Zeeman splitting and Overhauser shifts are plotted for OD II for the valence holes (left panel) and conduction electrons (right panel) }
\end{figure}

Now we turn to estimating the values of the hole hyperfine constants $C_1$ and $C_2$ given by Eq.~\eqref{eq:C1C2} and $C_{\rm eff}$ given by Eq.~\eqref{Ceff}.  A hole in the dot experiences interaction with both Ga and As nuclei, therefore the hyperfine constants should be averaged over the spin polarization of two nuclei species. Assuming equal spin polarizations of the Ga and As nuclei, we obtain
\begin{equation}
\label{est:0}
\frac{C_{1,2}}{A} = \frac{C_{1,2}^{({\rm Ga})} + C_{1,2}^{({\rm As})}}{A^{({\rm Ga})} + A^{({\rm As})}}\:.
\end{equation}
The values of the electron hyperfine constants are well established~\cite{Urbaszek:2013a}: $A^{({\rm Ga})} = 42$~$\mu$eV and $A^{({\rm As})} = 46$~$\mu$eV. The hole hyperfine coupling constants are still debated in the literature. The detailed fitting of the present experimental data and the data of Ref.~\cite{Chekhovich:2013a} should be carried out in a consistent manner and it is beyond the scope of the present work. Here, to illustrate that such the fitting may result in an agreement we take, for instance, $\tilde M_p^{({\rm Ga})}=0.045$, $\tilde M_d^{({\rm Ga})}=0.06$, $\tilde M_p^{({\rm As})}=0.08$, $\tilde M_d^{({\rm As})}=0$, which results in the somewhat smaller absolute value of the $C^{({\rm Ga})}/A^{({\rm Ga})}=-0.045$ and the larger value of the $C^{({\rm As})}/A^{({\rm As})}=0.19$ in the (001) QDs, which are still inside the error bounds of Ref.~\cite{Chekhovich:2013a}. For this choice of parameters and $g_{h2}>0$ we have the $|C_{\rm eff}/A| \approx 0.06$ for both trigonal QDs. This value is slightly smaller than the experimentally extracted ratio $|C_{\rm eff}/A| = 0.2\pm 0.1$ but, given the error margin of the current data and the data of Ref.~\cite{Chekhovich:2013a}, such an agreement can be considered as satisfactory. Further, the agreement may be improved by taking into account the orbital contribution to $C_2$ caused by the heavy-light hole mixing, Eq.~\eqref{C2lh}.

To provide a complete description the further experimental and theoretical studies are needed. Particularly, on experimental side element-sensitive measurements on the (111) QDs will provide an additional information on the signs and magnitudes for the ratio $C_{\rm eff}/A$ for Ga and As isotopes. This, together with the data of Ref.~\cite{Chekhovich:2013a}, would provide   a more complete picture of the hyperfine coupling for holes and establish stronger restrictions on the model parameters. As for the theory, a simultaneous fitting of data on (001) and (111) quantum dots taking into account also the orbital contribution to $C_2$ is challenging.



%

\end{document}